# DYNAMIC MODELLING OF COMBINED CYCLE POWER PLANT FOR LOAD FREQUENCY CONTROL WITH LARGE PENETRATION OF RENEABLE ENERGY

Songyao Jiang

Department of Electrical Engineering and Computer Science, Graduate School of University of Michigan
songyaoj@umich.edu

Supervisor: Associate Prof. KATO, Takeyoshi

Graduate School of Engineering, Nagoya University
tkato@nuee.nagoya-u.ac.jp

## ABSTRACT

As the concern about climate change and energy shortage grow stronger, the incorporation of renewable energy in the power system in the future is foreseeable. In a hybrid power system with a large penetration of PV generation, PV panel is regarded as a negative load in the power system. With the accurate prediction of PV output power, load frequency control could be done by controlling the thermal and hydro power plant in the system. Combined Cycle Power Plant is widely used because of its great advantages of fast response and high efficiency. This article is focusing on the mathematical modelling and analyzing of Combined Cycle Power Plant for the frequency control purpose in a model of hybrid system with large renewable energy generation.

## 1. BACKGROUND

### 1.1 FUTURE OF RENEWABLE ENERGY

Electricity is regard as a kind of clean energy without pollution in utilization. But nowadays, the process to generate electricity is not clean because thermal power plant which combust coal oil and natural gas still hold an important part of the generation. These kind of energy sources are called fossil fuel which are converted from the ancient dead organisms through a long natural process called decomposition. Because the process, millions of years, is significantly longer than any human activities, fossil fuels are generally considered to be non-renewable resources because. In addition, the consumption of fossil fuel always comes together with the emission of pollution to air and water system. Therefore, as concerns of the shortage and pollution of fossil fuel grows stronger, many countries encourage the development of renewable energy. It is expected that the renewable energy could completely replace the traditional fossil fuel and even nuclear in the future.

The World Wide Fund for Nature (WWF), which is an international non-governmental organization working on issues regarding environment, post a famous energy report in 2011. In this report, WWF advocates a 100% renewable energy future in 2050.[1] It analyse the demand and supply of energy and provide and approach to the utilization of renewable energy. It also points out the challenges ahead like equity, land and sea use, lifestyle, innovation and provide some possible solutions. Peoples are motivated by this report that the world is excited for the coming of a foreseeable renewable future. Figure 1 shows the percentage of different energy source from 2000 to 2050.

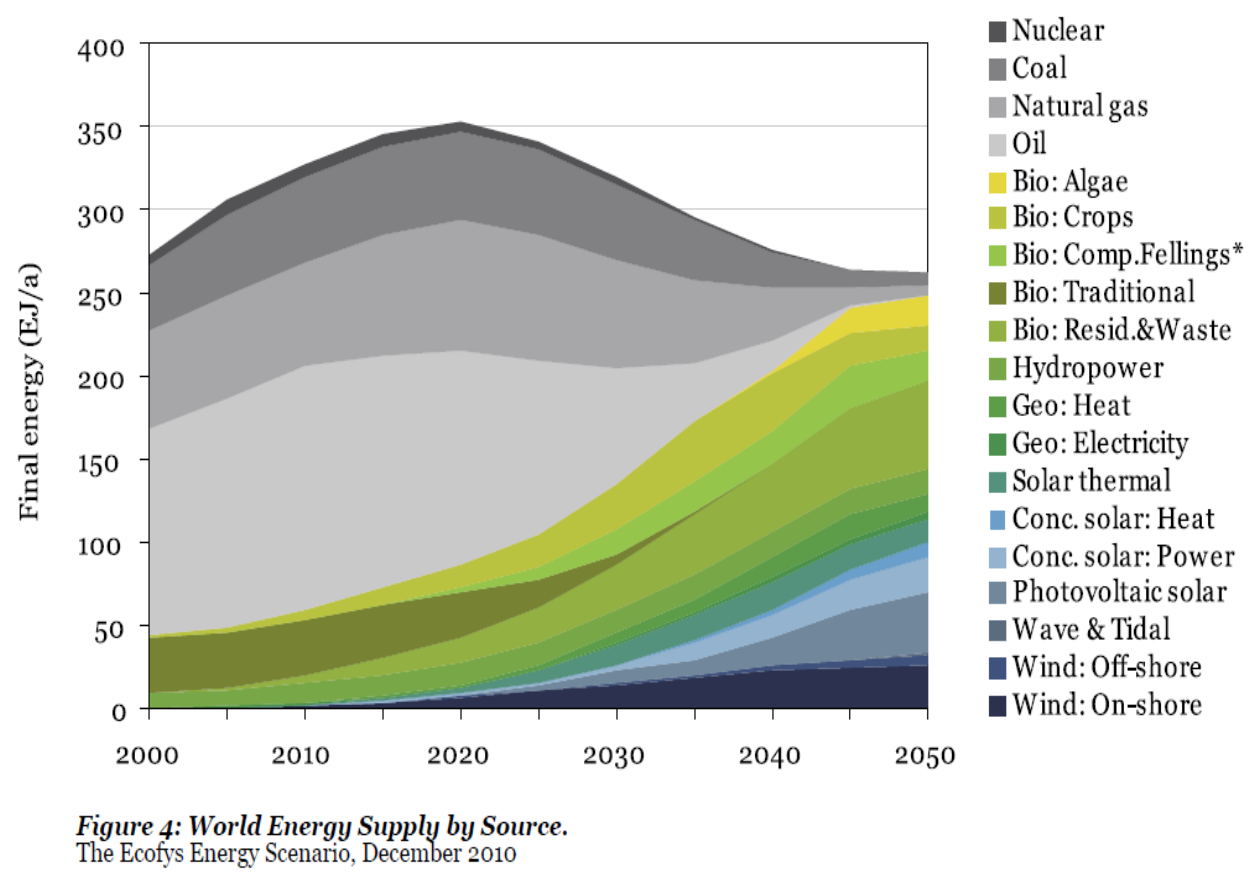

Fig.1 World Energy Supply by Source [1]

The figure shows that in the future renewable energy like biomass, hydropower and solar power will play a major role in the energy consumption. Some countries, for example Germany and Australia are reported targeting switch to the 100% renewable energy in 2050. Because of the consideration of climate changes and energy shortage, replacing traditional fossil and nuclear energy with renewable energy has become a worldwide trend.

## 1.2 RENEWABLES IN POWER SYSTEM

Currently, most of the electricity generation relies on the combustion of fossil fuel. Figure 2 is a pie chart presenting the percentage of power generation by different type of sources for the first four months of 2014. This chart is delivered from the data provided by US energy department.

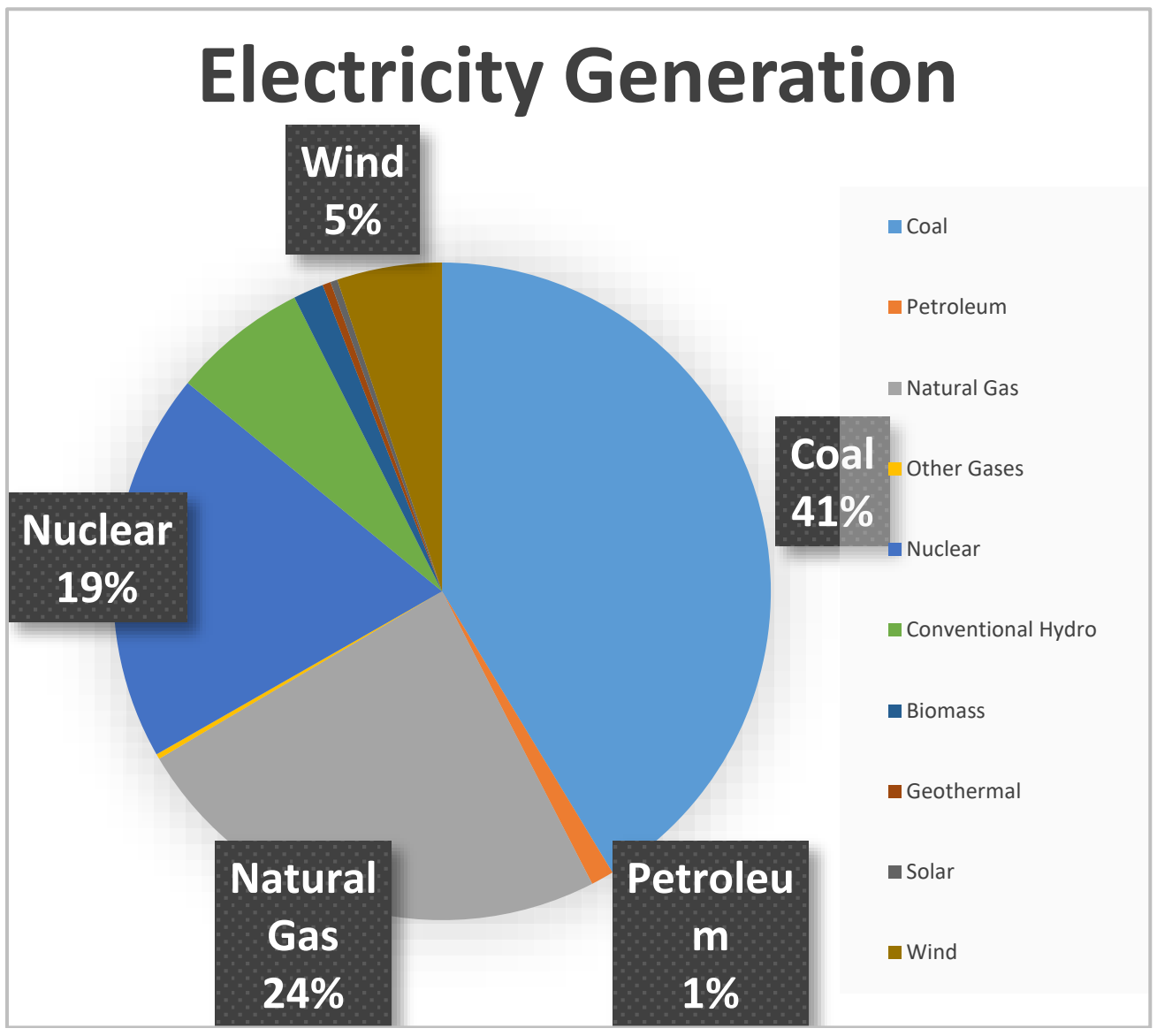

Fig. 2 Electricity Generation by Sources [2]

As shown in the figure, coal and natural gas forms 65% of the total power generation and renewables only forms less than 15% of the generation now. Wind energy and traditional hydropower are two of the most mature renewables in the world. However, wind farm is considered as a project that take large area of land to operate. Wind farm must choose the area where wind is relatively strong and sustaining over the year. This kind of area is always far from the place where the power is majorly used, for example big cities and large factories. Solar power is considered to be one of the most important resources in the future with large potential. As we know, most of the energy on earth comes from the sun. If a very small part of that power could be utilized, human will never suffer the shortage of energy. WWF made an assumption in their energy report that if 0.3% of the Sahara Desert was a concentrated solar plant, it would power all of the Europe. Of course, this idea sounds too idealist. In fact, just like wind farm, PV farm also requires a significantly large area of land to install. This feature increase the cost of PV power and prevents PV farm to be developed further. Recently the development of the idea Micro-grid and Distributed Generation incorporate large usage of solar panel. Unlike the traditional power system that generation and load are separated to be far from each other, Distributed Generation is an ideal that generation is installed at the usage site. PV panel, wind turbine, small natural gas generator and fuel cell are the major generation plant in distributed generation. For example, in a district of living houses, PV panels could be installed on every roof of the buildings, which makes usage of the sunlight that not used currently and solve the issue that PV panels need large area to install. Large penetration of renewable energy generation is definitely practical and foreseeable in the future.

As renewable energy generation depends mostly on the condition of the weather, for example PV panel depends on the amount of sunlight that received and wind turbine depends on the speed of wind, it is hard to regulate the generation of renewables comparing to traditional generation. While controlling the power output of renewables, the response is much slower than traditional generation with petroleum and natural gas. For a hybrid power system with renewables and traditional fossil fuel power plant, the generation of PV panels is often regarded as negative load in the system. The responsibility of balancing load and generation relies on the control of traditional generation in the system. The load and generation control is discussed in detail in next section. The techniques to predict PV output power is still under development. PV output power could be predicted by imaging processing using satellite images or ground based camera. Many of the students in my lab are currently working on the forecasting of PV power now. Renewable energy generation will definitely play an important role in power system in the future.

## 2. AGC OF HYBRID POWER SYSTEM

As known to all, electricity cannot be stored in significant amount. Therefore, the dynamic balance between generation and demand is one of the most important issue in power system. It is required that load and generation are balanced moment by moment. Load is the usage of electricity in the grid, so in general, it could not be controlled. Therefore, the generation of power plant should be controlled and adjusted frequently to balance the system. Nowadays, this process is controlled automatically and so called *Automatic Generation Control* (AGC).

In the process of balancing the load and generation, it is difficult to directly measure the load. Instead, as the grid is using AC current nowadays, system frequency is considered to be a very useful indicator of the power mismatch. Power mismatch will have an influence on the system frequency. If the generation is larger than demand, the generation will converted to kinetic energy and therefore increase the frequency of the system. The amount of increase in system frequency depends on the total storage and total inertia of the system. If the generation in the power system could not meet the demand, the kinetic energy will convert to electricity to meet the demand of load. Therefore, if the frequency becomes lower than nominal value, which is 50Hz or 60Hz, it indicates that the generation needs to be increased. If the frequency is higher than the nominal value, the generation needs to be decreased. Continues adjustment need to be implemented in the system to keep the frequency stable. It is because that frequency, as an aspect of system stability, is important for many applications on the grid. Large deviation and fluctuation in frequency will cause potential damage to some precise instruments.

### 2.1 TURBINE-GOVERNOR CONTROL

There are different types of AGC. The first one is called *Turbine-Governor Control*. The kinetic energy talk above is stored in the turbine generator units in the power system. The turbine flows the Newton's second law, which is

$$J\alpha = T_m - T_e \tag{1}$$

When the load suddenly increase, the current will also increase. This will lead the electrical torque *Te* of each turbine unit to increase to supply the increased load. From the equation (2), it could be seen that the acceleration α then becomes negative. A negative acceleration will cause the drop of speed of the turbine units. Because the system frequency is proportional to the speed of the generating synchronous machine, the frequency also drops.

The Turbine-Governor Control act as a proportional control to compensate the changes in load. It follows the equation:

$$\Delta p_m = \Delta p_{ref} - \frac{1}{R}\Delta f \tag{2}$$

where $\Delta p_m$ is the change in turbine $\Delta f$ is the frequency deviation, $\Delta p_{ref}$ is the change of the setting of reference power. R is called the droop coefficient. Figure 3 shows a Turbine-Governor Control example.

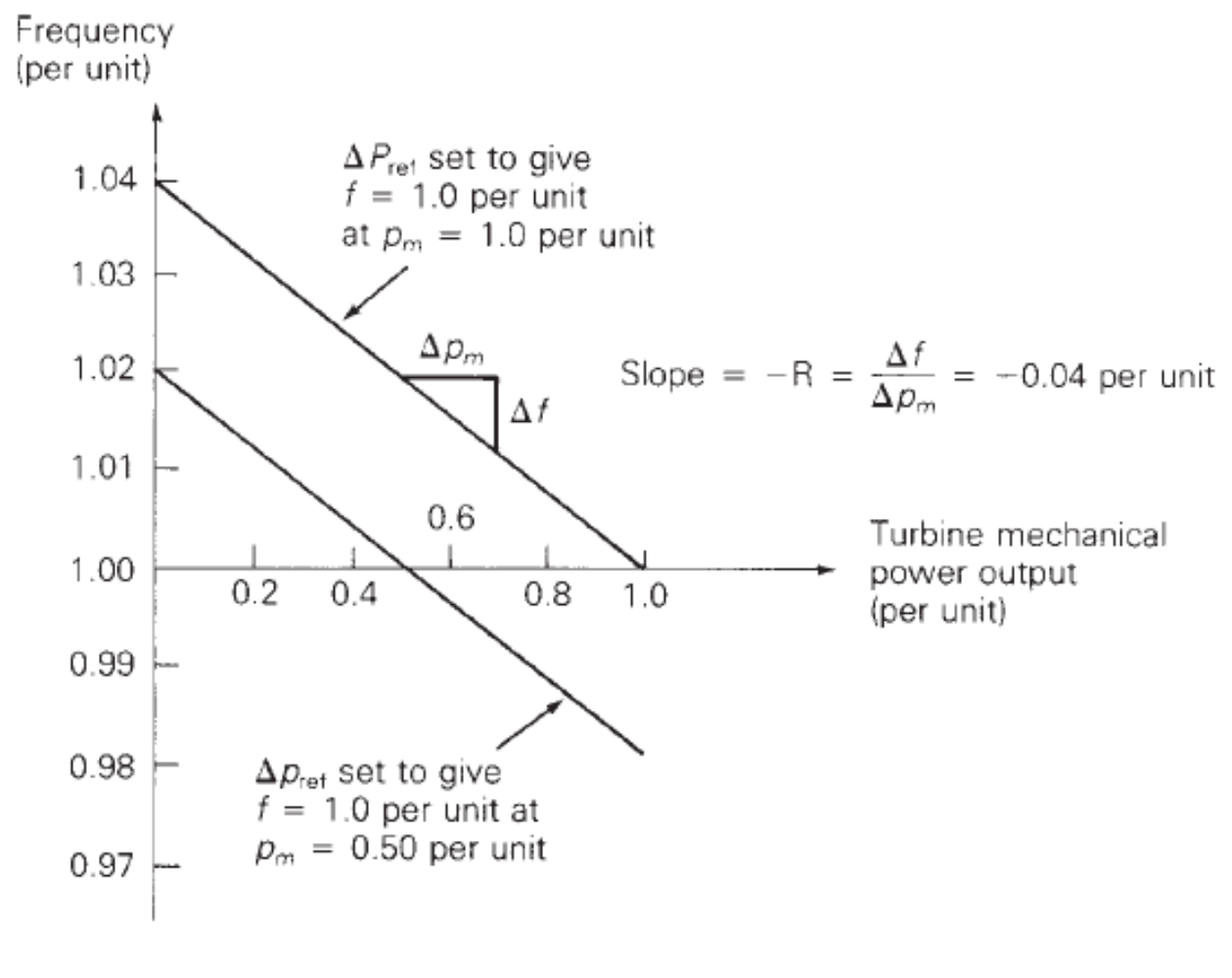

Fig. 3 Turbine-Governor Control [X]

### 2.2 LOAD-FREQUENCY CONTROL

Because the Turbine-Governor Control is proportional control, there exists a steady state error in frequency. The second type of AGC, which is *load frequency control* (LFC), could further management the output of generators and reset the frequency to its nominal value. Considering a control area in the power system with interconnected tie-lines, the objective of LFC is to absorb its own load changes in the area, regulating the frequency to nominal value and maintaining the power flow in tie-lines as scheduled. There is an important factor in LFC called *area control error* (ACE). It could be defined using the following equation:

$$ACE = \Delta p_{tie} + B_f \Delta f \tag{3}$$

where $\Delta f$ is the frequency deviation and $\Delta p_{tie}$ is the deviation of net tie-line power flow. $\Delta p_{ref}$ is the change of the setting of reference power. $B_f$ is called the *frequency bias constant*.

LFC works together with Turbine-Governor that $\Delta p_{\mathrm{ref}i}$ could be calculated as the integral of ACE, which is,

$$\Delta p_{\mathrm{ref}\,i} = -K_i \int \mathrm{ACE}\, dt \tag{4}$$

With the integral part in this control process, the frequency deviation will be continuously affect the output of generators to maintain the frequency at its nominal value.

## 3. COMBINED CYCLE POWER PLANT

### 3.1 INTRODUCTION OF CCGT

The gas turbine engine is a complex assembly of different components such as compressors, turbines, combustion chambers, etc., designed on the basis of thermodynamic laws [1]. It is a type of heat engines that absorb the heat of energy source, converting it into mechanical energy, which drives electrical generators. Combined cycle gas turbine is based on the design of single cycle gas turbine. As shown in Figure 4. In single cycle gas turbine, when the gas turbine is operated, air is filtered and then drawn through the inlet of the compressor. The fuel, which is natural gas in this case, is fed into the combustor together with compressed air and combusted. The resulted high-temperature and high-pressure gas is fed into the gas turbine and drives its shaft. Then the mixed gas is exhausted. However, this exhaust gas is still low in its Entropy which means much energy is wasted without utilization. Therefore, it limits the efficiency of single cycle gas turbine, which is about 34% in net Carnot thermo-dynamic efficiency.

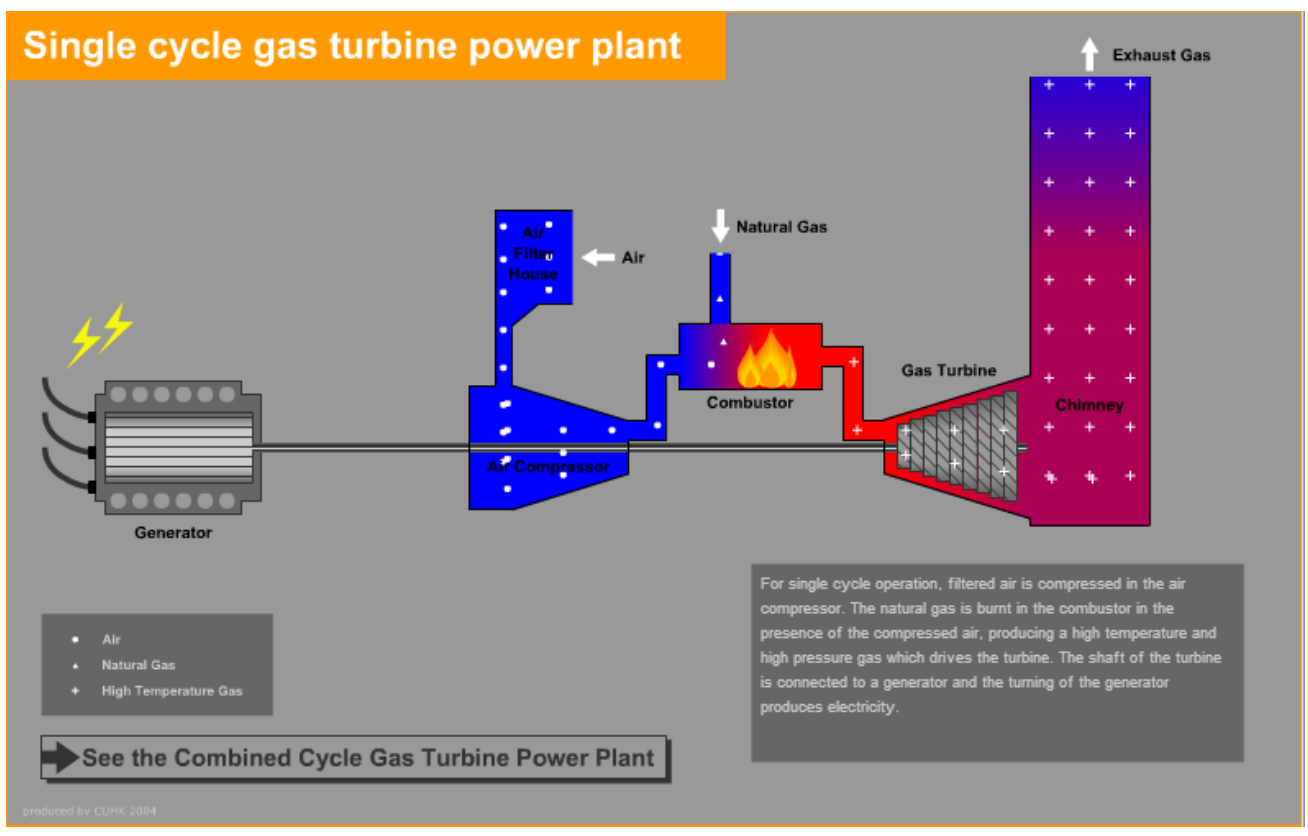

Fig. 4 Single cycle gas turbine power plant

By adding a subsequent heat engine, it may extract energy from the waste heat of the exhaust gas, which could improve the overall efficiency of the plant. A commonly used combination is adding a steam turbine after the cycle of gas turbine. The remaining heat in exhaust gas generate steam in a special device called *heat recovery steam generator* (HRSG), as shown in Figure 5. The steam is fed into the steam turbine and produce additional mechanical power and therefore electricity. The steam is recycled through condenser and fed into HRSG again. This process could probably improve 50 – 60 percent of the overall efficiency. This is called a Combined Cycle Gas Turbine. It could achieve a thermal efficiency of around 60%. Because of its high efficiency and fast response, many new gas power plants around the world are of this kind.

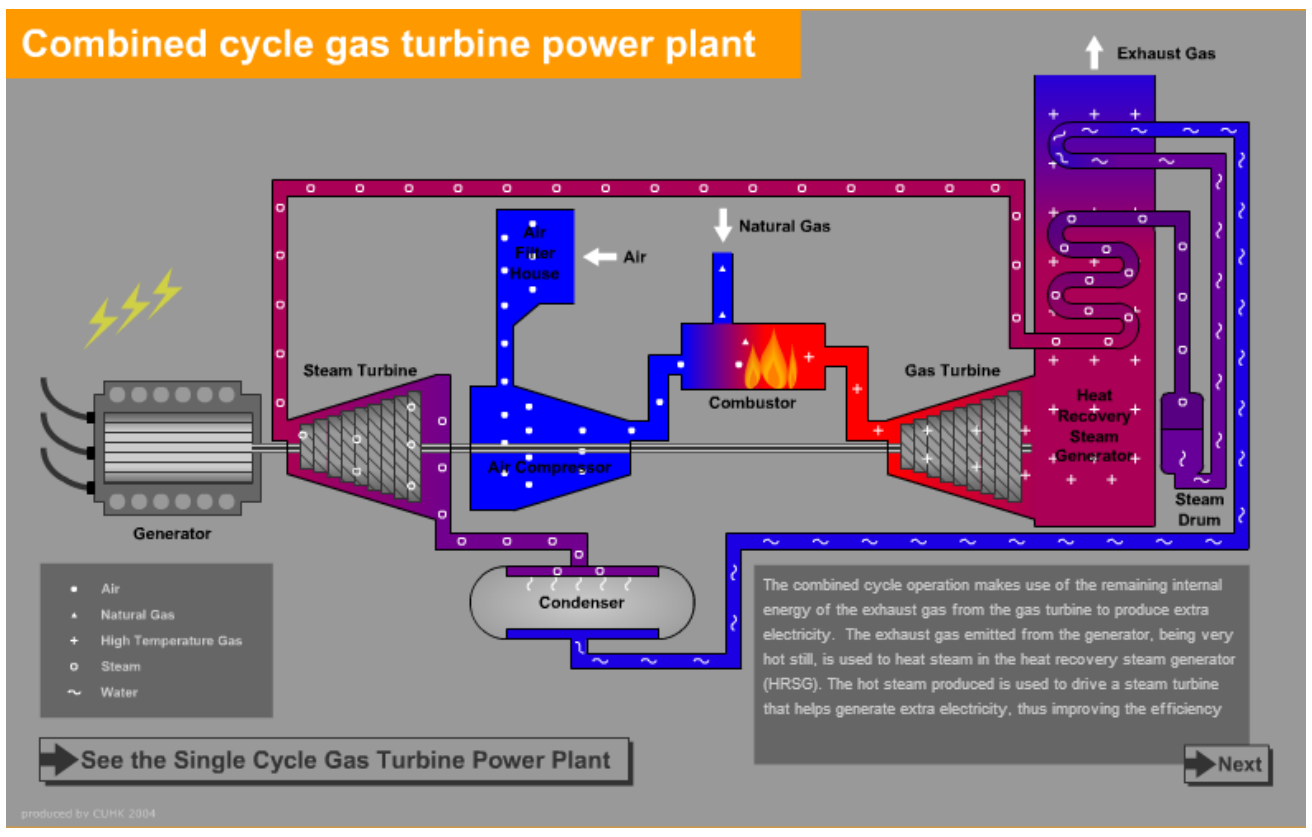

Fig. 5 Combined cycle gas turbine power plant

## 3.2 MATHEMATICAL MODELLING & LITERATURE SURVEY

To simulate load frequency control, the dynamic characteristics of combined cycle power plant is crucial. As a heat engine, thermal effect plays the most important role in modelling combined cycle power plant. There are two major points of control in this kind of power plant, which are air flow and fuel flow. The air flow into the compressor is controlled by inlet of the compressor while the fuel flow into the compressor is controlled by the fuel valve. Air and fuel together determine the temperature and pressure of the resulted gas output from the combustor, which determines the power output of the gas turbine and the temperature of the exhaust gas. To model the dynamic performance of combined cycle, it is better to use the single cycle gas turbine as a basic reference. There is a famous model of gas turbine which is called Rowen's Model. [4]

### 3.2.1 ROWEN'S MODEL OF SINGLE CYCLE GAS TURBINE.

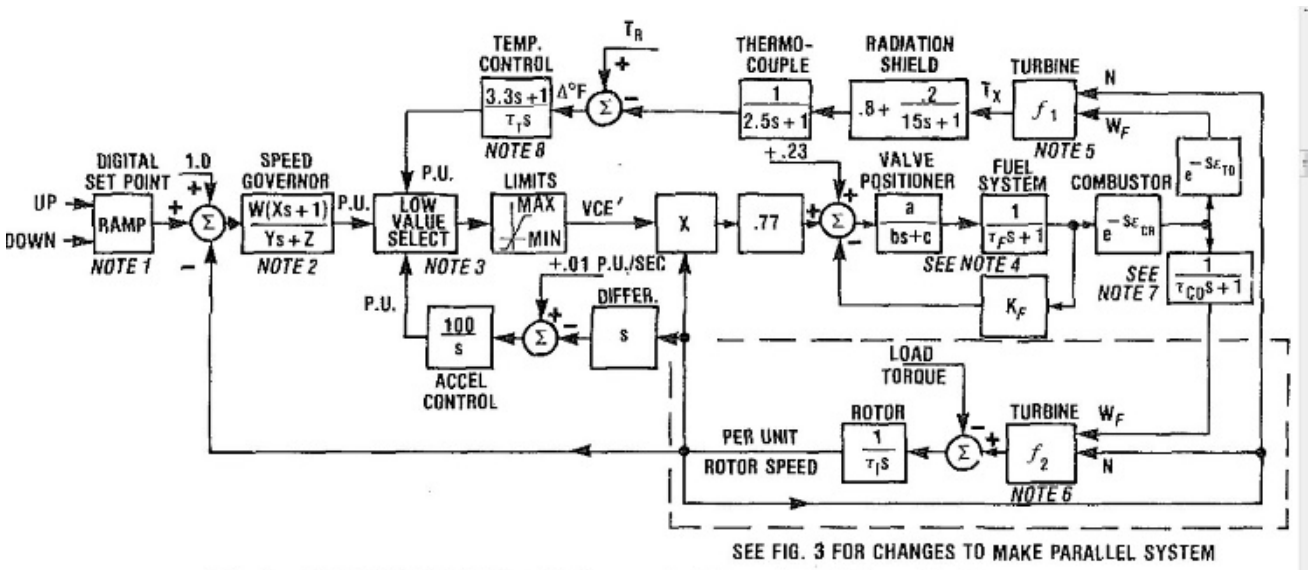

Fig. 1 Simplified single-shaft gas turbine simulation block diagram for isolated operation[3]

Fig. 6 Rowen's model of SCGT [4]

Rowen provided a very detailed explanation of his model. In this model, it could be found that thermal flow and control both play critical roles. As shown in Figure 6, there are three different types of control in his model. The first one is *speed control*. The measured speed from the rotor dynamics is subtracted from the speed set point, which is 1 *per unit* (pu). The error signal is fed into the speed governor and generate speed control signal. The second one is the *temperature control* circuit. The exhaust temperature is calculated from the gas flow and rotating speed of the turbine. Then the temperature value is transmitted through the blocks of thermocouple and radiation shield. This process simulates the measurement of the exhaust temperature that the resulted temperature is regarded as the measured temperature by devices. The temperature controller controls the temperature of the exhaust gas and set the exhaust temperature at reference value to prevent the gas turbine from overheat. It could also improve the efficiency of the gas turbine. The third one is *acceleration control*. This controller is used to prevent the turbine from accelerate or decelerate too fast. The low value selector compared three control signals and use the lower one to determine the operating condition of the gas turbine. This control signal is regarded as the fuel signal which determine the fuel flow into the combustor. The time constants of fuel system and combustor delay are also considered and included. The resulted gas flow into the turbine together with the system frequency are used to calculate the turbine output force and exhaust temperature. The output force is fed into the block which stands for the rotor inertia. Together with the load torque, regarding the inertia as an integrator, the rotor speed could be calculated easily.

### 3.2.2 ZHANG'S MODEL OF COMBINED CYCLE GAS TURBINE

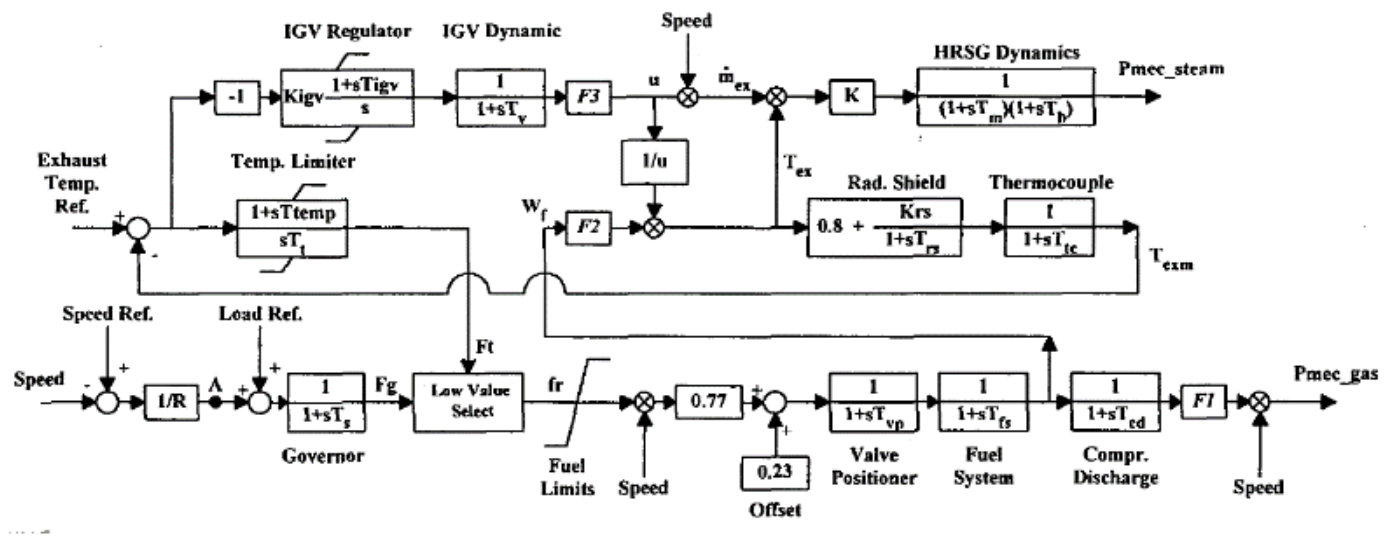

Fig. 2. Combined cycle plant dynamic model

Fig. 7 Zhang's model of CCGT [5]

For combined cycle gas turbine, because of the existence of HRSG and steam turbine. The mathematical model is more complicated than the model of single cycle gas turbine. But the Rowen's model of SCGT could be a great reference. Zhang's model is based on the model build by Rowen. As explained in Zhang's article, the time constant of steam turbine, which is the second cycle in CCGT, is much slower than the constant of gas turbine, which is the first cycle in CCGT. In fact, the exhaust temperature in CCGT is controlled at certain fixed value to maximize the efficiency of steam turbine. Therefore, as a result, the dynamics of steam turbine have little influence on the overall dynamics of CCGT. As shown in Figure 7, Zhang's model is basically very similar to Rowen's model. Acceleration control is eliminated in Zhang's model, IGV control is added to control the temperature of exhaust gas to maximize the efficiency and the dynamics of HRSG and steam turbine is added. Steam flow is calculated as a function of turbine speed and a coefficient K is added to stand for the efficiency of steam turbine. The output power of gas turbine and steam turbine could be added together to stand for the total power output of CCGT.

### 3.2.3 MANTZARIS & VOURNAS'S MODEL OF COMBINED CYCLE GAS TURBINE

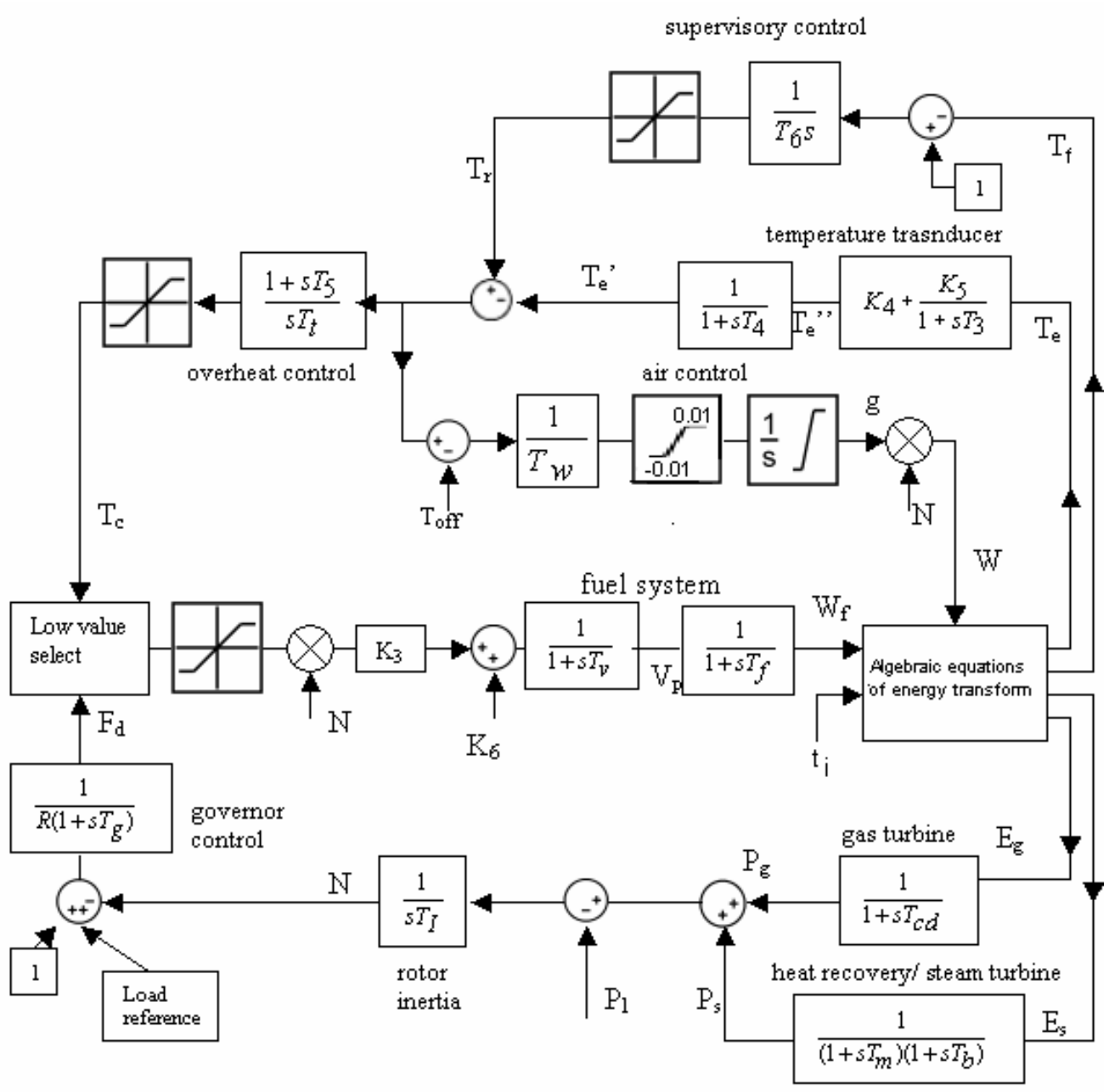

Fig. 8 Mantzaris & Vournas's model of CCGT [6]

Mantzaris & Vournas develop their own model and incorporate the advantages of Zhang's model by making their own modification. As shown in Figure 8, all the mathematical calculation is integrated in a block called *Algebraic Equation of Energy Transform*. This model has a clear flow of signals that the key control variables, which are air flow W and fuel flow Wf are fed into the algebraic block and calculate gas flow and steam flow into the turbines together with the temperature of exhaust gas. There is an additional control circuit in this model called *supervisory control*. It takes the gas turbine inlet temperature into consideration that the inlet temperature will have an influence on reference temperature which is used to regulate exhaust temperature to maintain the plant at maximum efficiency.

Because of the logic structure of this model is clear and the parameters are given in detail. I rebuild my model according to this one. The following's are the equations used in this model [7]:

From the adiabatic compression equation the following relation holds, where $x$ is the ratio of input-output temperatures for isentropic compression:

$$x = \frac{t_{d,is}}{t_i} = (P_r)^{\frac{\gamma-1}{\gamma}} \quad (1)$$

In (1) $P_r$ is the actual compressor ratio. For nominal airflow (W=1pu), this is equal to the nominal ration $P_{r0}$. When airflow is different from nominal (W≠1), the actual compression ratio is

$$P_r = P_{r0}W \quad (2)$$

and therefore:

$$x = (P_{r0}W)^{\frac{\gamma-1}{\gamma}} \quad (3)$$

From the definition of compressor efficiency:

$$\eta_c = \frac{t_{d,is} - t_i}{t_d - t_i} \quad (4)$$

The gas turbine inlet temperature depends on the fuel to air ratio (assuming that air is always in excess). The temperature rises with the fuel injection $W_f$ and decreases with airflow W.

From the energy balance equation in the combustion chamber, the following normalized equation results:

$$W\frac{t_f - t_d}{t_{f0} - t_{d0}} = W_f \quad (6)$$

or:

$$t_f = t_d + (t_{f0} - t_{d0})\frac{W_f}{W} \quad (7)$$

Similar to (4), the gas turbine efficiency is given by

$$\eta_t = \frac{t_f - t_e}{t_f - t_{e,is}} \quad (8)$$

For the adiabatic expansion, noting from (3) that the right hand side is the same as in the compression (the mass that enters the compressor is the same with the one in the output of the gas turbine) we have:

$$x = \frac{t_f}{t_{e,is}} \quad (9)$$

from which we obtain for the actual exhaust temperature similarly to (5):

$$t_e = t_f[1 - (1 - \frac{1}{x})\eta_t] \quad (10)$$

$$E_g = K_0[(t_f - t_e) - (t_d - t_i)]W \quad (11)$$

The thermal power absorbed by the heat exchanger of the recovery boiler is proportional to airflow and exhaust temperature.

$$E_s = K_1 t_e W \quad (12)$$

| | Parameter | Value |
|---|---|---|
| ti0 | Ambient temperature (K) | 303 |
| td0 | Nominal compressors discharge temperature (C) | 390 |
| tf0 | Nominal gas turbine inlet temperature (C) | 1085 |
| te0 | Nominal exhaust temperature(C) | 532 |
| Pr0 | Nominal compressor pressure ratio | 11.5 |
| γ | Ratio of specific heat (Cp/Cv) | 1.4 |
| ηc | Compressor efficiency | 0.85 |
| ηt | Turbine efficiency | 0.85 |
| K0 | Gas turbine output coefficient (1/K) | 0.00303 |
| K1 | Steam turbine output coefficient (1/K) | 0.000428 |
| R | Speed governor regulation | 0.04 |
| Tg | Governor time constant (s) | 0.05 |
| K4 | Gain of radiation shield | 0.8 |
| K5 | Gain of radiation shield II | 0.2 |
| T3 | Time constant of radiation shield(s) | 15 |
| T4 | Time constant of thermocouple(s) | 2.5 |
| T5 | Time constant of temperature control (overheat) (s) | 3.3 |
| Tt | Temperature control (overheat) integration rate (s) | 0.4699 |
| Tc max | Temperature control upper limit | 1.1 |
| Tc min | Temperature control lower limit | 0 |
| Fd max | Fuel control upper limit | 1.5 |
| Fd min | Fuel control lower limit | 0 |
| K3 | Fuel valve lower limit | 0.23 |
| Tv | Valve positioner time constant (s) | 0.05 |
| TF | Fuel system time constant (s) | 0.4 |
| T6 | Time constant of Tf control (s) | 60 |
| Tw | Time constant of air control (s) | 0.4699 |
| Tcd | Gas turbine time constant (s) | 0.2 |
| Tm | Steam turbine time constant (s) | 5 |
| Tb | Heat recovery boiler time constant (s) | 20 |
| TI | Turbine rotor inertia constant (s) | 18.5 |
| Toff | Temperature offset | 0.01 |

Table 1: Parameters used in M&V's Model of CCGT [6]

## 3.3 MODIFICATION & SIMPLIFICATION

I rebuilt the model according to Mantzaris & Vournas's model using Simulink. The model is shown in Figure 9.

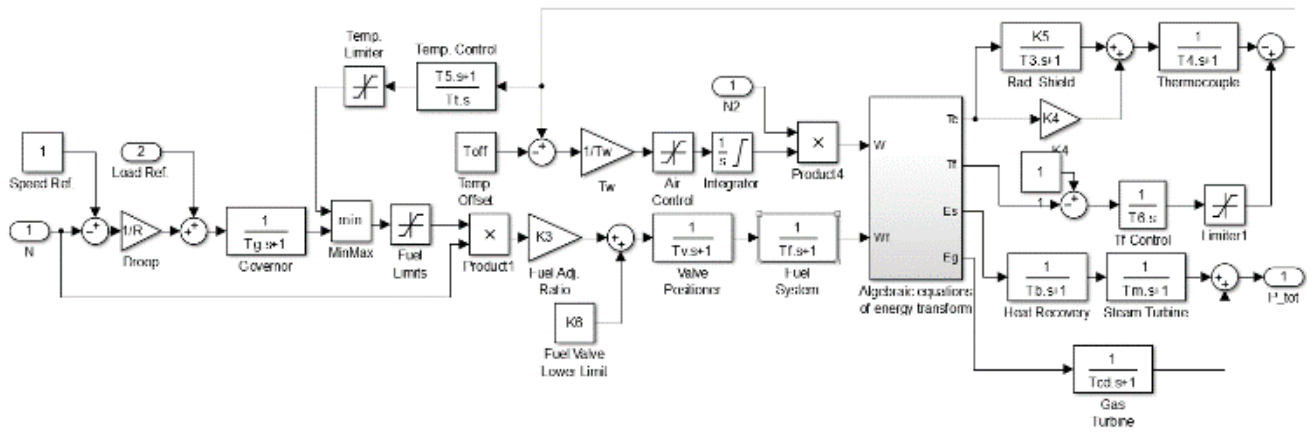

Fig. 9 Model of CCGT in Simulink

There are two input of this sub system, which are frequency and load reference, and one output, which is the overall output of the Combined Cycle Power Plant. In this project, the major concern is the regulation of frequency by adjusting the power output of plants in its normal operating point, which means the speed deviation is small. Therefore, the dynamics of the path from frequency deviation to power output is the most important path in our model. The above model is too complicated for this project because of the consideration about the thermal transform and control paths. The above model must be simplified before implemented.

According to the explanation in Rowen's article. The product of the turbine speed with air flow and fuel flow is not necessary when the speed deviation is small. Therefore, in our application, this product could be eliminated. Furthermore, the supervisory control is not necessary. It is because that the ambient temperature, which is the air inlet temperature is not given in our application. This temperature is considered to be constant which would have no influence on the reference temperature. The simplified version of CCGT model is shown in Figure 10.

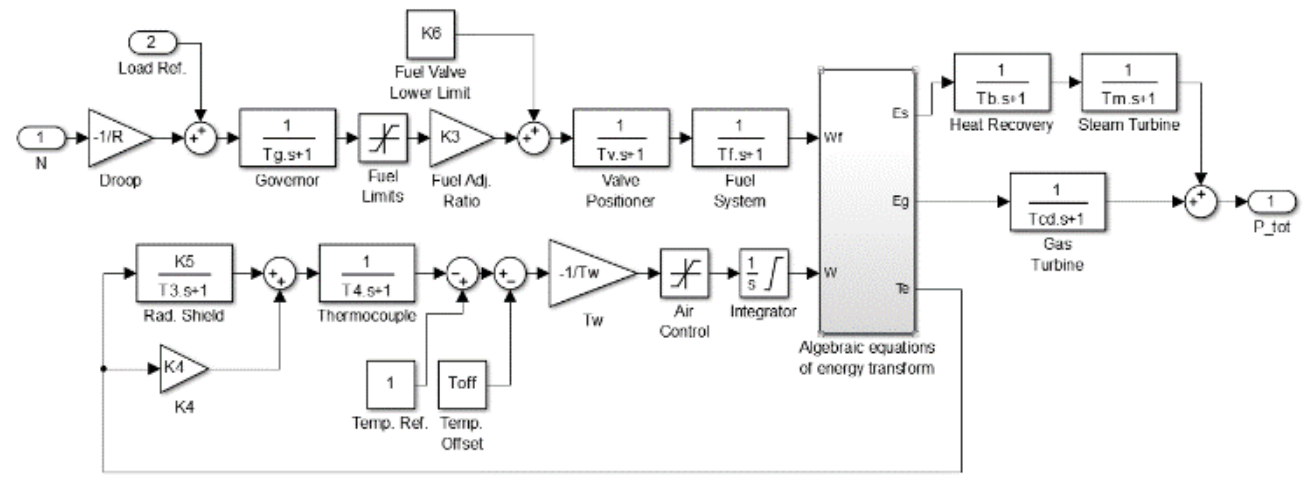

Fig. 10 Simplified CCGT Ver.1

The logic relationship is clear in this model. The two points of control, which are air flow W and fuel flow Wf, are decoupled comparing to the previous model. However, this model still seems to be complicated than expected. After careful consideration, I found that the air control loop is used for maintain the exhaust temperature at desired value to maximize the efficiency of the system. In my application, efficiency is not concerned, so we could make an assumption that the air flow is always at its nominal value, which is 1 pu. In this situation, the control loop of air flow could be completely deleted. In such case, the exhaust temperature will not be fed back and controlled. The further simplified version of model is shown in Figure 11 below.

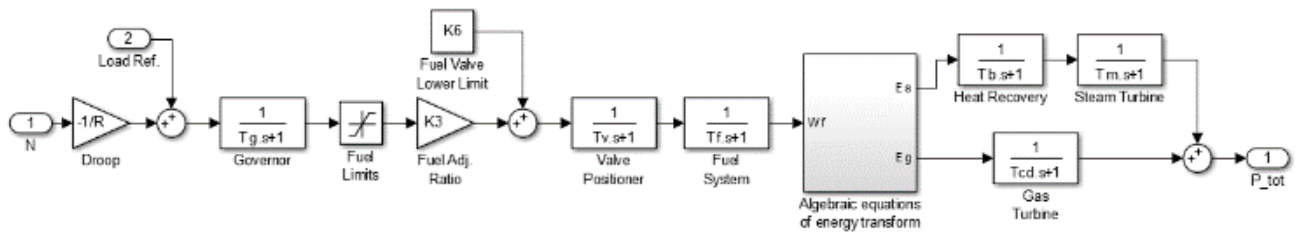

Fig. 11 Simplified CCGT Model Ver. 2

As shown in this model, there is a single loop through which the frequency deviation and power output is connected with many time constants and delays. This version of CCGT model is simplified enough and meets the requirements of this project.

# 4 SIMULATION AND RESULTS

## 4.1 SIMPLE TEST CIRCUIT

After modelling and simplified the dynamic behaviour of CCGT. Simple test must be implemented to examine the assumption I made in simplifying the model. Figure 12 shows a simple test circuit I built in Simulink to test the dynamic behaviour of the model after the sudden change of load.

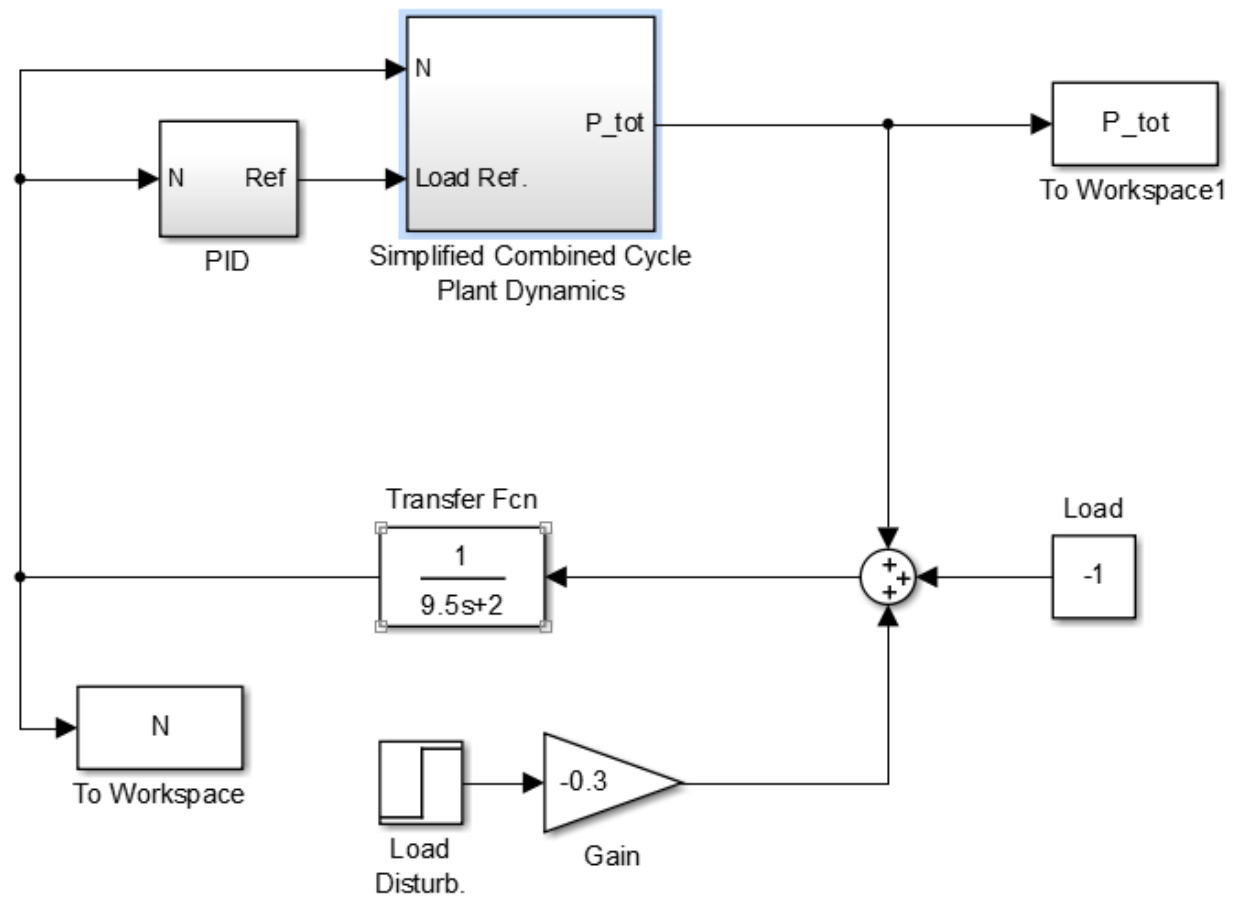

Fig. 12 Simple Test Circuit in Simulink

There is a constant load of 1 pu in the circuit and the load suddenly change to 1.3 pu at 200s using a step function and a gain of 0.3. The output power of CCGT is added to the total load of the system. The resulted mismatch is fed into a transfer function which stands for the total system inertia. The parameters of the inertia is get from the inertia of Matsumoto's model. The output of the inertia dynamics is the speed deviation. This deviation is regulated externally by a simple PI controller which is shown in Figure 13.

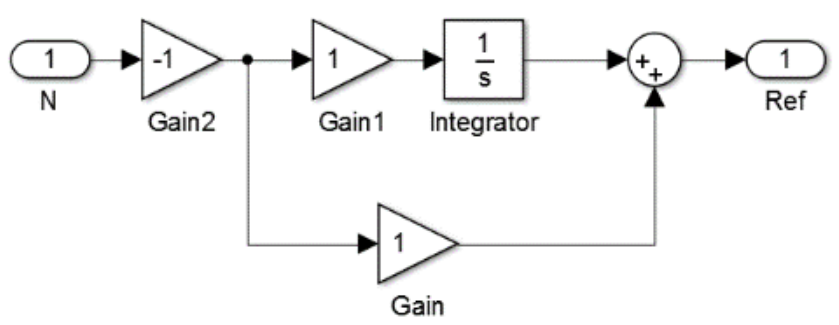

Fig. 13 Simple PI Controller

The control signal output is fed into CCGT model as the load reference signal which is used as an integral control to reset the frequency at its nominal value.

By comparing the dynamic behaviour of the model before and after simplification, the assumption I made during simplification could be therefore verified. Figure 14 shows the result of this simulation. It shows the power output and frequency change of the CCGT. The red line stand for the model before simplification and the blue line stand for the model after simplification. After the sudden change in load at 200s, the frequency suddenly drop by 0.02. Then the controllers, which are both Turbine Governor Control and LFC worked and got the frequency back. Then the output of the CCGT becomes 1.3 pu, which is equal to the total load. It means that the generation and demand are balanced again. There is almost no difference between the model before and after simplification. There is only tiny difference at the start. Starting behaviour is not a concern in this project. Therefore, according to this simple test, the simplified model of CCGT is thought to be valid and sound.

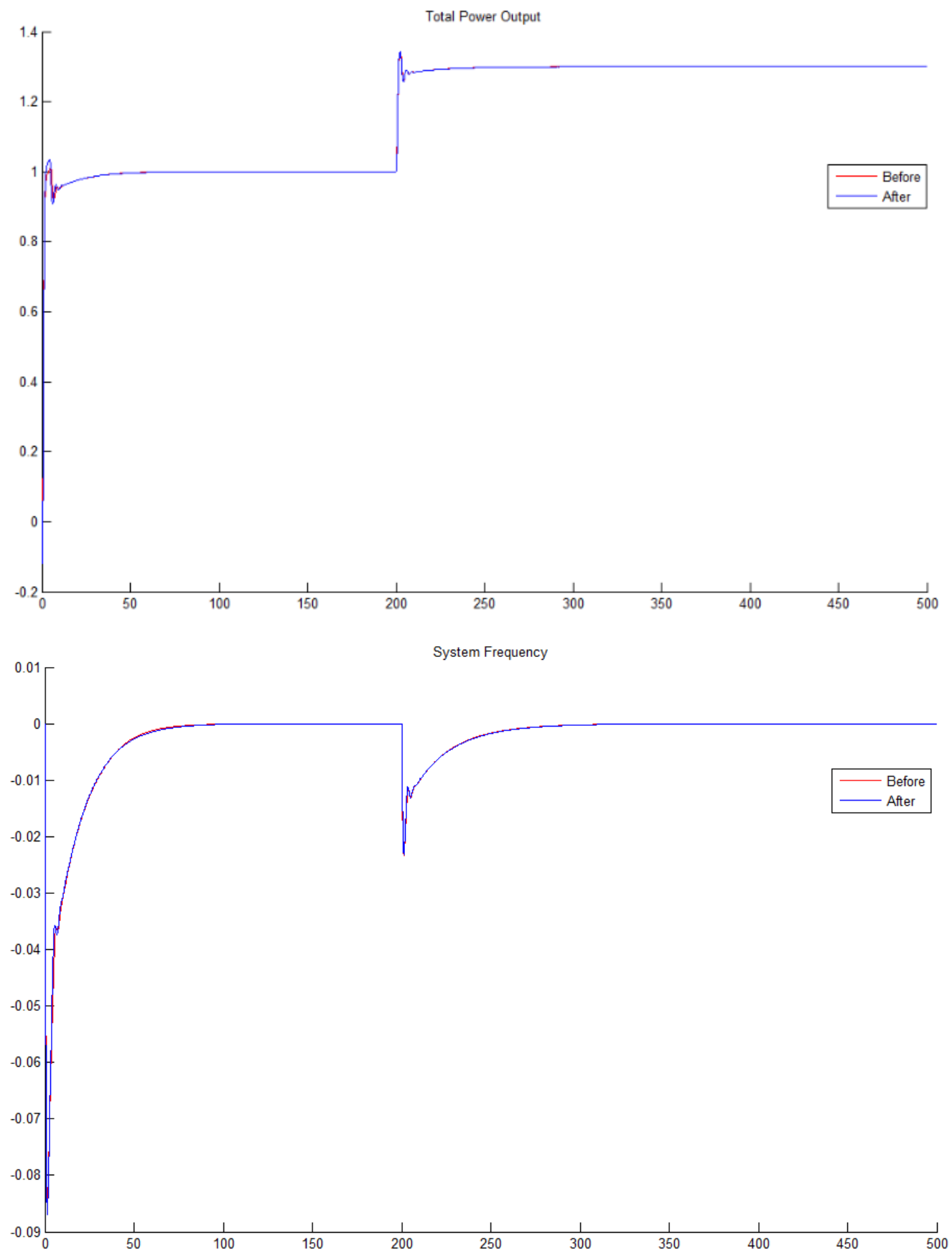

Fig. 14 Comparison of Simulated Results

### 4.2 IMPLEMENT ON MATSUMOTO'S MODEL

Mr. Matsumoto tested my model of CCGT together with his model. However, in your model, there is energy loss in the simulation block to calculate the exhaust gas temperature and turbine input gas temperature. Therefore, my CCGT model initially did not work as of it in his model. He made a modification to my model that the output gain of the abovementioned block was adjusted to cancel the energy loss. After that, he replaced the steam turbine model to my CCGT model. At this time my CCGT model works fine with his system. As a result, the frequency deviation was reduced because of the quicker response of CCGT than steam turbine.

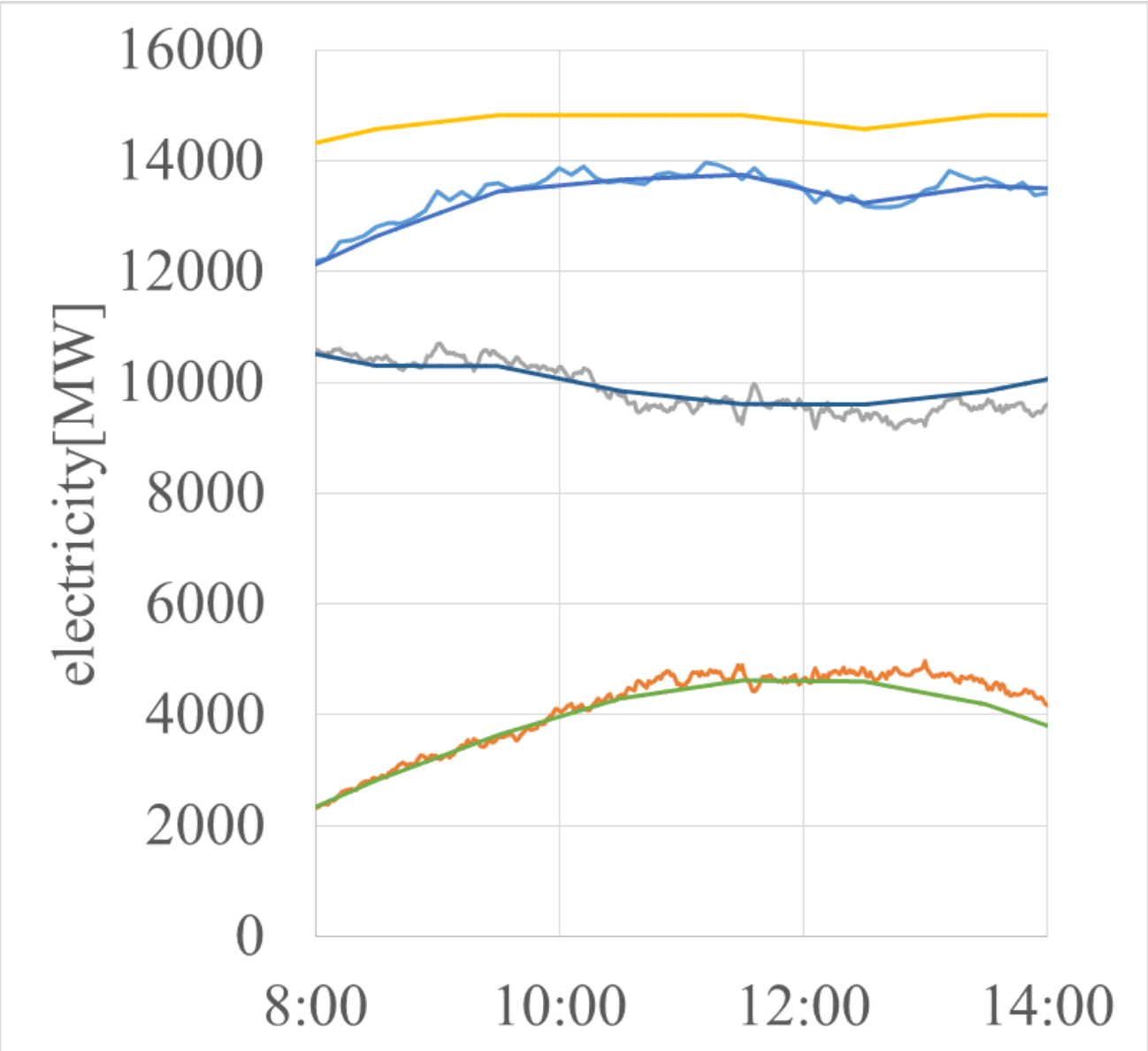

Fig. 15 Total Power of Matsumoto's Hybrid System

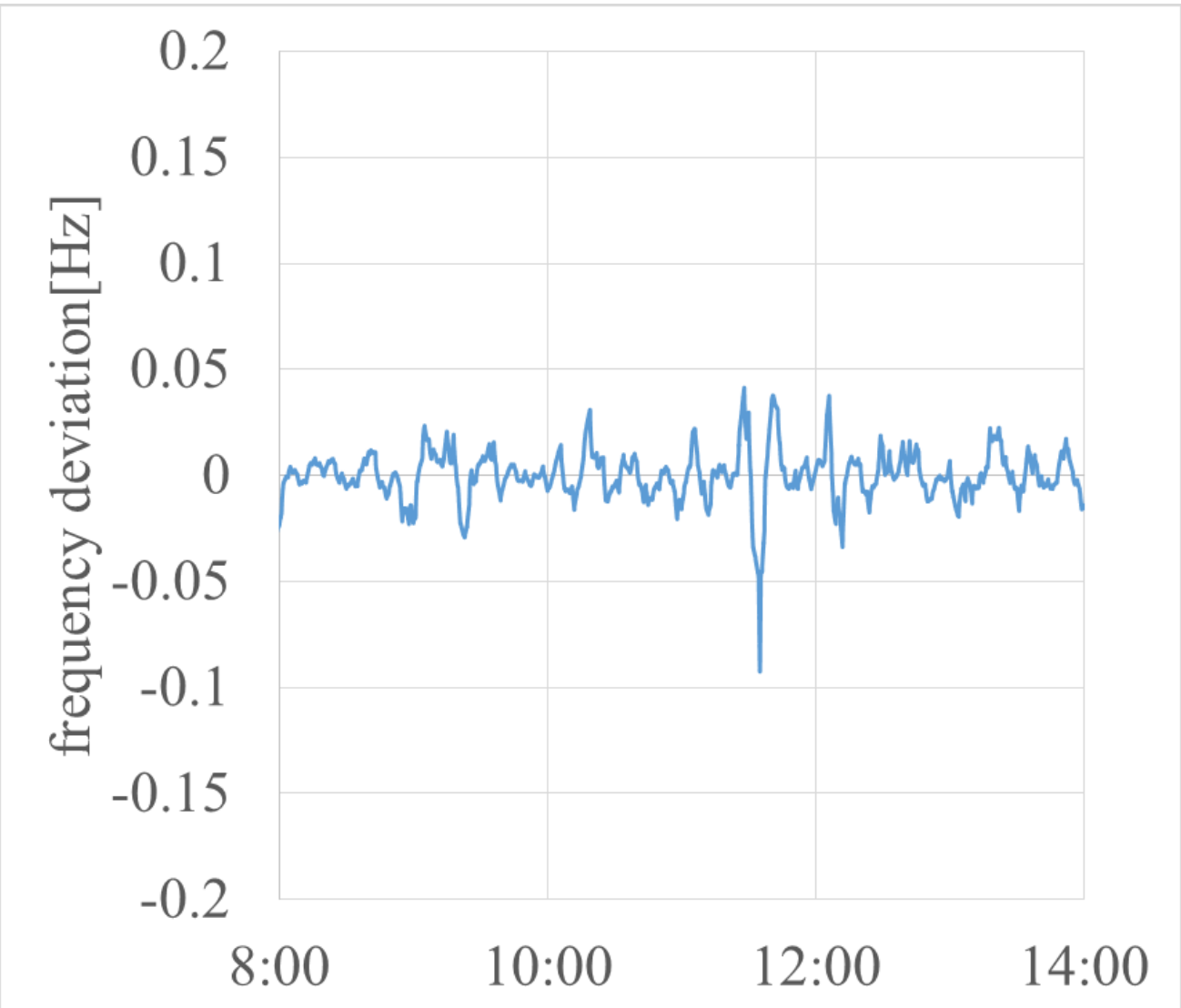

Fig. 16 Frequency Deviation Using the Original Steam Turbine

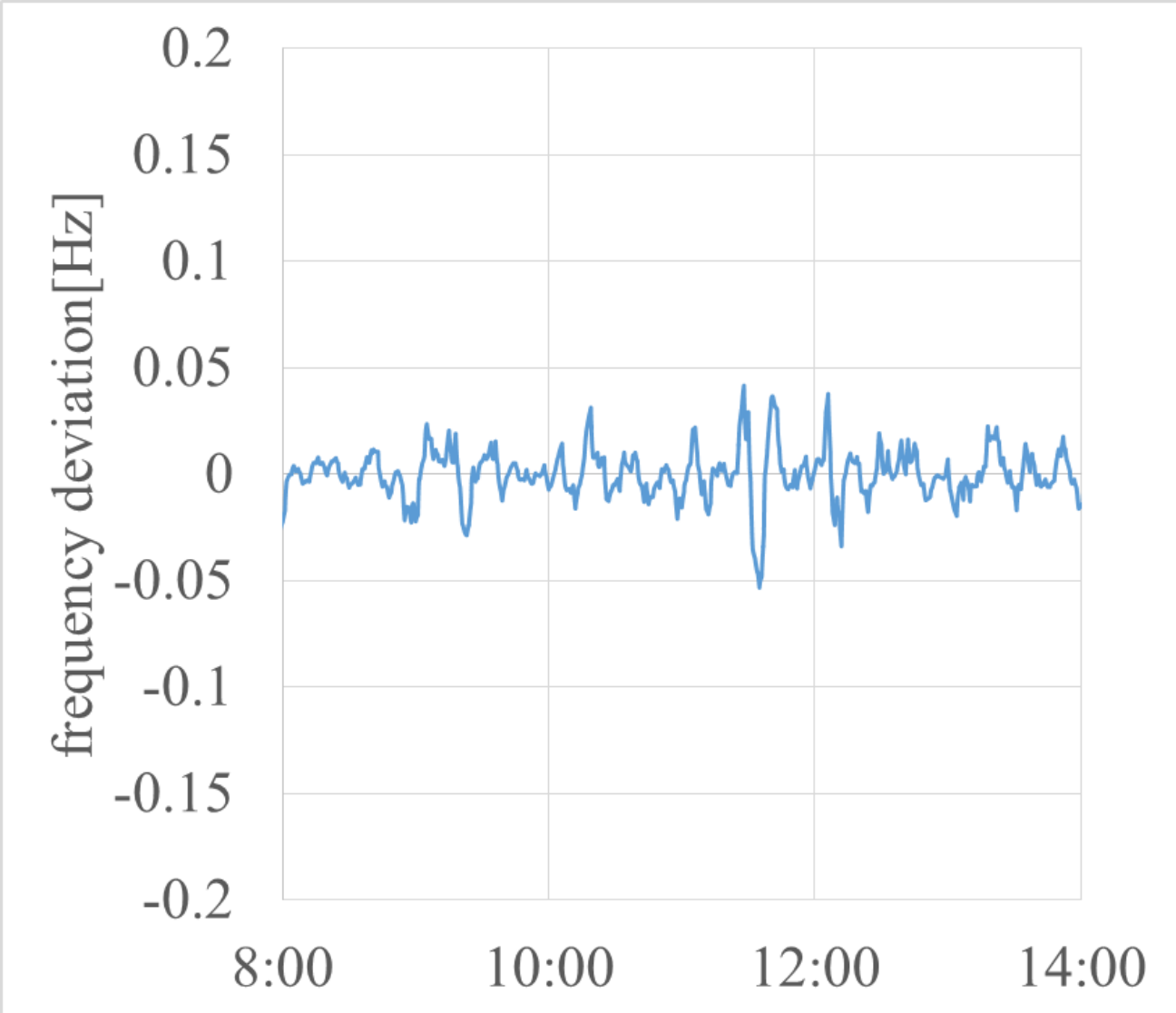

Fig. 17 Frequency Deviation Using the CCGT

## ACKNOWLEDGEMENTS

I would like to express my deepest appreciation to all those who provided me the possibility to complete this report. A special gratitude I give Supervisor, Associate Professor. Kato, whose contribution in stimulating suggestions and encouragement, helped me to coordinate my project especially in simplifying the model.

Furthermore I would also like to acknowledge with much appreciation the crucial role of Mr. Matsumoto, who explained me the theories in his model and help me simulate my simplified model with his hybrid system.